\documentclass[aps,prc,preprint,showpacs,showkeys]{revtex4}
\usepackage{amssymb}
\usepackage{amsmath}
\usepackage{graphicx}
\usepackage{bm}
\usepackage{epsf}

\begin{document}

\title{Quark-meson coupling model for antikaon condensation in neutron star
matter with strong magnetic fields}
\author{P. Yue}
\affiliation{Department of Physics, Nankai University, Tianjin 300071, China}
\author{H. Shen}
\email{songtc@nankai.edu.cn}
\affiliation{Department of Physics, Nankai University, Tianjin 300071, China}

\begin{abstract}
We study the effects of strong magnetic fields on antikaon condensation in
neutron star matter using the quark-meson coupling (QMC) model. The QMC
model describes a nuclear many-body system as nonoverlapping MIT bags in
which quarks interact through the self-consistent exchange of scalar and vector
mesons in the mean-field approximation. It is found that the presence of
strong magnetic fields alters the threshold density of antikaon condensation
significantly. The onset of $K^-$ condensation stronger depends on the magnetic
field strength, and it even shifts beyond the threshold
of $\bar K^0$ condensation for sufficiently strong magnetic fields.
In the presence of strong magnetic fields, the equation of state (EOS) becomes
stiffer in comparison with the field-free case. The softening of the EOS by
antikaon condensation also depends on the magnetic field strength, and it
becomes less pronounced with increasing magnetic field strength.
The results of the QMC model are compared with those obtained in a relativistic
mean-field (RMF) model, and we find there are quantitative differences
between the results of the QMC and RMF models.
\end{abstract}

\maketitle

%\email{yuep@mail.nankai.edu.cn}

%\bibliographystyle{prsty2}

% 11.30.Rd Chiral symmetries
% 12.40.Yx Hadron mass models and calculations
% 21.30.Fe Forces in hadronic systems and effective interactions
% 21.60.-n Nuclear structure models and methods
% 21.65.+f Nuclear matter
% 21.80.+a Hypernuclei
% 24.10.Jv Relativistic models
% 24.85.+p Quarks, gluons, and QCD in nuclei and nuclear processes
% 26.60.+c Nuclear matter aspects of neutron stars
% 97.60.Jd Neutron stars (see also 26.60.+c Nuclear matter aspects)
% 98.35.Eg Electric and magnetic fields
% 95.30.Cq Elementary particle processes
% 95.30.Tg Thermodynamic processes, conduction, convection, equations of state
% 95.30.Wi Dust processes (condensation, evaporation, sputtering, mantle growth, etc.)

%\newpage
%%%%%%%%%%%%%%%%%%%%%%%%%%%%%%%%%%%%%%%%%%%%%%%%%%%%%%%%%%%%%%%%%%%%%

\section{Introduction}

\label{introduction}

The study of dense stellar matter in the presence of strong magnetic fields
is of great interest in nuclear astrophysics. Observations of ordinary radio
pulsars indicate that they possess surface magnetic fields of the order of
$10^{12}$ G~\cite{HL06}. Recent surveys of soft-gamma repeaters (SGRs) and
anomalous X-ray pulsars (AXPs) imply that the surface magnetic field of young
neutron stars could be of order $10^{14}$-$10^{15}$ G~\cite{TD96}. The magnetic
field strength may vary significantly from the surface to the center in neutron
stars. So far, there is no direct observational evidence for the internal
magnetic fields of the star, while it may reach $10^{18}$ G, as estimated in
some theoretical works~\cite{Latt00,Latt01,Mao03}.
Motivated by a possible existence of strong magnetic fields in neutron stars,
theoretical studies on the effects of extremely large fields on dense matter
have been carried out by many authors~\cite{HL06,Latt00,Mao03,CBP97,Mao06,YS06},
and the inclusion of hyperons and boson condensation has also been
investigated~\cite{Latt02,Pion01,Kaon02}.
There have been some works that investigate the effects of strong magnetic
fields on neutron star properties~\cite{Latt01,Mao03}.
In Ref.~\cite{Latt01}, the authors
studied static neutron stars with poloidal magnetic fields and a simple class
of electric current distributions, and they found that the maximum mass among
these static configurations with magnetic fields is noticeably larger than the
maximum mass attainable with uniform rotation and no magnetic field.
In Ref.~\cite{Mao03}, the authors assumed the magnetic field varies from
the surface to the center in neutron stars, and they found that the maximum
mass of neutron stars could substantially increase if the strongest possible
magnetic fields existed in the center of neutron stars.

It is believed that the density in the interior of neutron stars is
extremely high, and additional degrees of freedom such as hyperons,
kaons, and even quarks may occur in the core of neutron
stars~\cite{PR00,PRC96,PRC99,Panda04}. Recently, much attention has been
paid to the kaon/antikaon condensations based on various
models~\cite{PBW00,BB01,GS99,KPE95,PREPL00,GLZ03,Panda05}. In general,
the presence of antikaon condensation tends to soften the equation of
state (EOS) at high density and lower the maximum mass of neutron stars. It is
interesting to investigate the influence of strong magnetic fields on antikaon
condensation. In Ref.~\cite{Kaon02}, the condensation of negatively charged
$K^{-}$ under the influence of strong magnetic fields has been studied within
a relativistic mean-field (RMF) model. It was found that the threshold
of $K^{-}$ condensation shifts to higher density in the presence
of strong magnetic fields and the EOS becomes stiffer. These qualitative
features are expected to persist in other models.

In this article, we study the effects of strong magnetic fields on the
condensations of negatively charged $K^{-}$ and neutral $\bar{K}^{0}$ in
neutron star matter using the quark-meson coupling (QMC) model. The QMC
model was originally proposed in the article by Guichon~\cite{QMC88},
in which the quark degrees of freedom are explicitly taken into account
and a nuclear many-body system is described as a collection of nonoverlapping
MIT bags interacting through the self-consistent exchange of meson mean fields.
The QMC model has been extended and applied to various problems of nuclear
matter and finite nuclei with reasonable
success~\cite{QMC96,QMC98,QMC98h,QMC07}.
Furthermore, the model has also been used to investigate the properties of
neutron stars with the inclusion of hyperons, quarks,
and $K^{-}$ condensation~\cite{PRC99,Panda04,Panda05}. In the present work,
both nucleons and antikaons are described as MIT bags that interact through
the self-consistent exchange of isoscalar scalar and vector mesons ($\sigma$
and $\omega$) and isovector vector meson ($\rho$) in the mean-field
approximation. These exchanged mesons couple directly to the confined quarks
inside the bags. In contrast to the RMF model, the quark structure plays a
crucial role in the QMC model, and the basic coupling constants are defined
at the quark level. In Ref.~\cite{QMC98k}, the in-medium kaon and antikaon
properties have been studied in the QMC model. In Ref.~\cite{Panda05}, the
consequences of including $K^{-}$ condensation on the EOS of neutron star
matter and related compact star properties have been investigated using the
QMC model. Next, we examine the effects of strong magnetic fields on $K^{-}$
and $\bar{K}^{0}$ condensations occurring in neutron star matter within the
QMC model.

The aim of the present article is to investigate the influence of strong magnetic
fields on antikaon condensation that may occur in the core of massive stars.
For antikaon condensation in the field-free case, there have been many
discussions in the literature after the paper by Kaplan and Nelson~\cite{KN86}
who pointed out the possibility of the existence of $K^{-}$ condensation in
dense nuclear matter. In general, the chemical potential of antikaons in dense
matter decreases with increasing density due to their interaction with nucleons.
As a consequence, the ground state of hadronic matter at high density might
contain antikaon condensation. The appearance of antikaon condensation can
soften the EOS of neutron star matter and lower the maximum mass of neutron
stars~\cite{PBW00,Panda05}.
However, it is known that the surface magnetic field of young neutron
stars can be of order $10^{14}$-$10^{15}$ G~\cite{TD96}, whereas the internal
magnetic fields of the star may reach $10^{18}$ G, as estimated in some
theoretical works~\cite{Latt00,Latt01,Mao03}. The possible existence of strong
magnetic fields in neutron stars motivates us to study the effects of extremely
strong magnetic fields on dense matter. It has been found that the composition
and the EOS of neutron star matter can be significantly affected by strong
magnetic fields~\cite{Latt00,Mao06,YS06}, while the maximum mass of neutron
stars might substantially increase if the strongest possible magnetic
fields existed in the center of neutron stars~\cite{Latt01,Mao03}.
Because antikaon condensation can be important in the understanding of neutron
stars, an investigation of the influence of strong magnetic fields on antikaon
condensation would be of interest for the study of compact stars in astrophysics.

This article is arranged as follows. In Sec.~\ref{sec:2}, we briefly describe
the QMC model for neutron star matter with antikaon condensation in the
presence of strong magnetic fields. In Sec.~\ref{sec:3}, we show and discuss
the numerical results in the QMC model and make a systematic comparison with
the results of the RMF model. Section~\ref{sec:4} is devoted to a summary.

%%%%%%%%%%%%%%%%%%%%%%%%%%%%%%%%%%%%%%%%%%%%%%%%%%%%%%%%%%%%%%%%%%%%%

\section{ Formalism}

\label{sec:2}

We adopt the QMC model to describe neutron star
matter with antikaon condensation in the presence of strong magnetic fields.
In the QMC model, nucleons and antikaons are described as MIT bags that
interact through the self-consistent exchange of isoscalar scalar and vector
mesons ($\sigma$ and $\omega$) and isovector vector meson ($\rho$) in the
mean-field approximation. To perform the calculation for neutron star matter
in the presence of strong magnetic fields, we first study the properties of
nucleons and antikaons under the influence of external meson and
electromagnetic fields using the MIT bag model. These external fields are
in principle functions of position in the bag, which may cause a deformation
of the bag. For simplicity, we neglect the spatial variation of the fields
over the small bag volume and take the values at the center of the bag as
their average quantities~\cite{QMC96}. We note that the spherical bag
approximation may be violated in a superstrong magnetic field where the
deformation of the bag can be significant.

For nucleons described as spherical MIT bags with external meson and
electromagnetic fields, the up and down quarks inside the bag satisfy
the Dirac equation
\begin{equation}
\left[ i\gamma _{\mu}\partial ^{\mu}-\left( m_{q}+g_{\sigma}^{q}\sigma
\right) -g_{\omega}^{q}\omega_{\mu}\gamma ^{\mu}
-g_{\rho}^{q}\tau_{3q}\rho_{3\mu}\gamma ^{\mu}
-\frac{e\left( 1+3\tau_{3q}\right)}{6}A_{\mu
}\gamma ^{\mu}\right] \psi _{q}=0,  \label{eq:quark}
\end{equation}
where $g_{\sigma}^{q}$, $g_{\omega}^{q}$, and $g_{\rho}^{q}$ are the
quark-meson coupling constants and $m_{q}$ is the current quark mass.
$\tau_{3q}$ is the third component of the Pauli matrices.
$\sigma$, $\omega_{\mu}$, $\rho_{3\mu}$, and $A_{\mu}$ are the values
of the meson and electromagnetic fields at the center of the bag.

The normalized ground state for a quark in the bag is given by
\begin{equation}
\psi _{q}(\mathbf{r},t)=\mathcal{N}_{q}\,e^{-i\epsilon _{q}t/R_{N}}\left(
\begin{array}{c}
j_{0}(x_{q}r/R_{N}) \\
i\,\beta_{q}\,\vec{\sigma}\cdot \hat{\mathbf{r}}\,j_{1}(x_{q}r/R_{N})%
\end{array}%
\right) \,{\frac{\chi _{q}}{\sqrt{4\pi}}},
\end{equation}
where
\begin{equation}
\beta_{q}=\sqrt{{\frac{\Omega_{q}-R_{N}\,m_{q}^{\ast}}
{\Omega_{q}\,+R_{N}\,m_{q}^{\ast}}}},
\end{equation}
\begin{equation}
\mathcal{N}_{q}^{-2}=2\,R_{N}^{3}\,j_{0}^{2}(x_{q})
\left[ \Omega_{q}(\Omega_{q}-1)+R_{N}\,m_{q}^{\ast}/2\right] /x_{q}^{2},
\end{equation}
with $\Omega_{q}=\sqrt{x_{q}^{2}+(R_{N}\,m_{q}^{\ast})^{2}}$ and $m_{q}^{\ast
}=m_{q}+g_{\sigma}^{q}\sigma$. $R_{N}$ is the bag radius of the nucleon,
and $\chi _{q}$ is the quark spinor. The boundary condition,
$j_{0}(x_{q})=\beta_{q}\,j_{1}(x_{q})$, at the bag surface determines
the eigenvalue $x_{q}$. The energy of a static nucleon bag consisting of three
ground-state quarks is then given by
\begin{equation}
E_{N}^{\mathrm{bag}}=3{\frac{\Omega_{q}}{R_{N}}}-{\frac{Z_{N}}{R_{N}}}+{%
\frac{4}{3}}\pi R_{N}^{3}B_{N},
\end{equation}%
where the parameter $Z_{N}$ accounts for various corrections including
zero-point motion and $B_{N}$ is the bag constant.
The effective nucleon mass is then taken to be
\begin{equation}
M_{N}^{\ast}=E_{N}^{\mathrm{bag}}.  \label{eq:mn}
\end{equation}%
The bag radius $R_{N}$ is determined by the equilibrium condition
$\partial M_{N}^{\ast}/ \partial R_{N}=0$.
In the present calculation, we take the
current quark mass $m_{q}=5.5$ MeV. The parameter $B_{N}^{1/4}=210.854$ MeV
and $Z_{N}=4.00506$, as given in Ref.~\cite{Panda04}, are determined by
reproducing the nucleon mass $M_{N}=939$ MeV and the bag radius $R_{N}=0.6$
fm in free space.

For antikaons, negatively charged $K^{-}$ and neutral $\bar{K}^{0}$,
we assume that they are described as MIT bags in the same way as
nucleons~\cite{Panda05,QMC98k}. The exchanged $\sigma$, $\omega$, and $\rho$
mesons are assumed to couple exclusively to the up and down quarks
(and antiquarks), not to the strange quark according to the
Okubo-Zweig-Iizuka (OZI) rule~\cite{Panda05}.
Hence the antiquarks, $\bar{u}$ in $K^{-}$ and $\bar{d}$ in $\bar{K}^{0}$,
and the $s$ quarks satisfy the Dirac equations%
\begin{equation}
\left[ i\gamma _{\mu}\partial ^{\mu}-\left( m_{q}+g_{\sigma}^{q}\sigma
\right) +g_{\omega}^{q}\omega_{\mu}\gamma ^{\mu}
+g_{\rho}^{q}\tau_{3q}\rho_{3\mu}\gamma ^{\mu}
+\frac{e\left( 1+3\tau_{3q}\right)}{6}A_{\mu}\gamma^{\mu}\right]
\psi_{\bar{q}}=0,  \label{eq:antiquark}
\end{equation}%
and%
\begin{equation}
\left[ i\gamma _{\mu}\partial ^{\mu}-m_{s}+\frac{e}{3}A_{\mu}\gamma ^{\mu
}\right] \psi _{s}=0.  \label{eq:squark}
\end{equation}%
Similarly, the effective mass of antikaons and kaons is given by%
\begin{equation}
m_{K}^{\ast}=\frac{\Omega_{q}+\Omega_{s}}{R_{K}}-\frac{Z_{K}}{R_{K}}+%
\frac{4}{3}\pi {R_{K}}^{3}B_{K},  \label{eq:mk}
\end{equation}%
where $\Omega_{q}=\sqrt{x_{q}^{2}+(R_{K}\,m_{q}^{\ast})^{2}}$
and $\Omega_{s}=\sqrt{x_{s}^{2}+(R_{K}\,m_{s})^{2}}$.
We take the strange quark mass
$m_{s}=150$ MeV and the bag constant $B_{K}=B_{N}$. The parameters
$Z_{K}=3.362$ and $R_{K}=0.457$ fm, as given in Ref.~\cite{Panda05}, are
determined from the kaon mass and the stability condition in free space.

To describe neutron star matter with antikaon condensation in the presence
of strong magnetic fields, we adopt the total Lagrangian density written at
the hadron level as the sum of nucleonic, kaonic, and leptonic parts,%
\begin{equation}
\mathcal{L}=\mathcal{L}_{N}+\mathcal{L}_{K}+\mathcal{L}_{l},  \label{eq:LT}
\end{equation}%
where
\begin{eqnarray}
\mathcal{L}_{N} &=&\sum_{b=n,p}\bar{\psi}_{b}\left[ i\gamma _{\mu}\partial
^{\mu}-q_{b}\gamma _{\mu}A^{\mu}-M_{N}^{\ast}-g_{\omega N}\gamma _{\mu
}\omega ^{\mu}-g_{\rho N}\gamma _{\mu}\tau_{iN}\rho_{i}^{\mu}-\frac{1}{2%
}\kappa _{b}\sigma_{\mu \nu}F^{\mu \nu}\right] \psi _{b}  \notag \\
&&+\frac{1}{2}\partial _{\mu}\sigma \partial ^{\mu}\sigma -\frac{1}{2}%
m_{\sigma}^{2}\sigma ^{2}-\frac{1}{4}W_{\mu \nu}W^{\mu \nu}+\frac{1}{2}%
m_{\omega}^{2}\omega_{\mu}\omega ^{\mu}  \notag \\
&&-\frac{1}{4}R_{i\mu \nu}R_{i}^{\mu \nu}
  +\frac{1}{2}m_{\rho}^{2}\rho_{i\mu}\rho_{i}^{\mu}
  -\frac{1}{4}F_{\mu \nu}F^{\mu \nu},  \label{eq:LN}
\end{eqnarray}%
\begin{equation}
\mathcal{L}_{K}={D}_{\mu}^{\ast}\bar{K}{\ D}^{\mu}K-m_{K}^{\ast 2}\bar{K} K,
\label{eq:LK}
\end{equation}%
\begin{equation}
\mathcal{L}_{l}=\sum_{l=e,\mu}\bar{\psi_{l}}\left[ i\gamma_{\mu}
\partial^{\mu}-q_{l}\gamma_{\mu}A^{\mu}-m_{l}\right] \psi_{l}.  \label{eq:LL}
\end{equation}%
Here $A^{\mu}=(0$, $0$, $Bx$, $0)$ refers to a constant external magnetic
field $B$ along the $z$ axis. The effective masses $M_{N}^{\ast}$ and
$m_{K}^{\ast}$ in Eqs.~(\ref{eq:mn}) and (\ref{eq:mk}) are
obtained at the quark level. The covariant derivative is defined as
$D_{\mu}=\partial_{\mu}+iq_{K}A_{\mu}+ig_{\omega K}{\omega_{\mu}}
+ig_{\rho K}\tau_{iK}\rho_{\mu}^{i}$. The isospin doublet for kaons is denoted
by $K\equiv \left( K^{+},K^{0}\right)$ and that for antikaons by
$\bar{K}\equiv \left( K^{-},\bar{K}^{0}\right)$. We note that the electric charges
of particles are $q_{e}=q_{\mu}=q_{K^{-}}=-e$, $q_{n}=q_{\bar{K}^{0}}=0$,
and $q_{p}$ $=e$. The anomalous magnetic moments of nucleons are
included with $\kappa_{p}=\mu_{N}\left( g_{p}/2-1\right) =1.7928\ \mu_{N}$
and $\kappa_{n}=\mu_{N}g_{n}/2=-1.9130\ \mu_{N}$, where $\mu_{N}$ is the
nuclear magneton. In the QMC model, the coupling constants at the hadron level
are related to the quark-meson coupling constants as $g_{\omega N}=3g_{\omega}^{q}$,
$g_{\omega K}=g_{\omega}^{q}$,
and $g_{\rho N}=g_{\rho K}=g_{\rho}^{q}$~\cite{Panda05,QMC98k}.
The quark-meson coupling constants
$g_{\sigma}^{q}=5.957$, $g_{\omega}^{q}=2.994$, and $g_{\rho}^{q}=4.325$ are
determined by fitting the saturation properties of nuclear matter~\cite{Panda04}.
The meson masses $m_{\sigma}=550$ MeV, $m_{\omega}=783$ MeV, and
$m_{\rho}=770$ MeV are used in the present calculation.

In the mean-field approximation, the meson field equations in the presence
of antikaon condensation and strong magnetic fields have the following forms:%
\begin{eqnarray}
m_{\sigma}^{2}\sigma &=&-\frac{\partial {M_{N}^{\ast}}}{\partial \sigma}
(\rho_{s}^{p}+\rho_{s}^{n})-\frac{\partial {m_{K}^{\ast}}}{\partial\sigma}
 \left( \frac{m_{K}^{\ast}}{\sqrt{m_{K}^{\ast 2}+|q_{K^{-}}|B}}\rho_{K^{-}}
+\rho_{\bar{K}^{0}}\right) ,  \label{eq:m1} \\
m_{\omega}^{2}\omega_{0} &=&g_{\omega N}\left( \rho_{v}^{p}+\rho_{v}^{n}\right)
 -g_{\omega K}\left( \rho_{K^{-}}+\rho_{\bar{K}^{0}}\right),  \label{eq:m2} \\
m_{\rho}^{2}\rho_{30} &=&g_{\rho N}\left( \rho_{v}^{p}-\rho_{v}^{n}\right)
 -g_{\rho K}\left( \rho_{K^{-}}-\rho_{\bar{K}^{0}}\right) . \label{eq:m3}
\end{eqnarray}%
The Dirac equations for nucleons and leptons are given by%
\begin{equation}
\left( {i\gamma_{\mu}\partial^{\mu}-q_{b}\gamma_{\mu}A^{\mu}
-M_{N}^{\ast}-g_{\omega N}\gamma^{0}\omega_{0}-g_{\rho N}\gamma^{0}
\tau_{3N}\rho_{30}-\frac{1}{2}\kappa_{b}\sigma_{\mu \nu}F^{\mu \nu}}\right)
\psi_{b}=0,  \label{eq:diracn}
\end{equation}%
\begin{equation}
\left( {i\gamma _{\mu}\partial ^{\mu}-q_{l}\gamma _{\mu}A^{\mu}-m_{l}}%
\right) \psi _{l}=0.  \label{eq:diracl}
\end{equation}%
The energy spectra for protons, neutrons, and leptons (electrons and muons)
are given by%
\begin{eqnarray}
E_{\nu ,s}^{p} &=&\sqrt{k_{z}^{2}+\left( \sqrt{{M_{N}^{\ast 2}+}2\nu {q_{p}B}%
}-s{\kappa _{p}B}\right) ^{2}}+g_{\omega N}\omega_{0}+g_{\rho N}{\rho_{30}}%
, \\
E_{s}^{n} &=&\sqrt{k_{z}^{2}+\left( \sqrt{{M_{N}^{\ast 2}}%
+k_{x}^{2}+k_{y}^{2}}-s{\kappa _{n}B}\right) ^{2}}+g_{\omega N}\omega_{0}
-g_{\rho N}{\rho_{30}}, \\
E_{\nu ,s}^{l} &=&\sqrt{k_{z}^{2}+m_{l}^{2}{+}2\nu \left\vert {q_{l}}%
\right\vert {B}},
\end{eqnarray}%
where $\nu =n+1/2-\text{sgn}\left( q_{i}\right) s/2=0,1,2,\ldots$ enumerates the
Landau levels of the fermion $i$ with electric charge $q_{i}$
($i=p,\ e,\ $ or $\mu$). The quantum number $s$ is $+1$ for spin up and $-1$
for spin down cases. The expressions of scalar and vector densities for protons
and neutrons are given by%
\begin{eqnarray}
\rho_{s}^{p} &=&\frac{{q_{p}B}M_{N}^{\ast}}{2\pi ^{2}}\sum_{\nu}\sum_{s}%
\frac{\sqrt{M_{N}^{\ast 2}+2\nu {q_{p}B}}-s\kappa _{p}B}{\sqrt{M_{N}^{\ast
2}+2\nu {q_{p}B}}}\ln \left\vert \frac{k_{f,\nu ,s}^{p}+E_{f}^{p}}{\sqrt{%
M_{N}^{\ast 2}+2\nu {q_{p}B}}-s\kappa _{p}B}\right\vert ,  \label{eq:rhosp}
\\
\rho_{v}^{p} &=&\frac{{q_{p}}B}{2\pi ^{2}}\sum_{\nu}\sum_{s}k_{f,\nu
,s}^{p},  \label{eq:rhovp} \\
\rho_{s}^{n} &=&\frac{M_{N}^{\ast}}{4\pi ^{2}}\sum_{s}\left[
k_{f,s}^{n}E_{f}^{n}-\left( M_{N}^{\ast}-s{\kappa _{n}B}\right) ^{2}\ln
\left\vert \frac{k_{f,s}^{n}+E_{f}^{n}}{M_{N}^{\ast}-s{\kappa _{n}B}}%
\right\vert \right] ,  \label{eq:rhosn} \\
\rho_{v}^{n} &=&\frac{1}{2\pi ^{2}}\sum_{s}\left\{ \frac{1}{3}k_{f,s}^{n3}-%
\frac{1}{2}s\kappa _{n}B\left[ \left( M_{N}^{\ast}-s{\kappa _{n}B}\right)
k_{f,s}^{n}\right. \right.  \notag \\
&&\left. \left. +E_{f}^{n2}\left( \arcsin \frac{M_{N}^{\ast}-s{\kappa _{n}B}%
}{E_{f}^{n}}-\frac{\pi}{2}\right) \right] \right\} ,  \label{eq:rhovn}
\end{eqnarray}%
where $k_{f,\nu ,s}^{p}$ and $k_{f,s}^{n}$ are the Fermi momenta of protons
and neutrons, which are related to the Fermi energies $E_{f}^{p}$ and
$E_{f}^{n}$ as
\begin{eqnarray}
{E_{f}^{p}}^{2} &=&k_{f,\nu ,s}^{p2}+\left( \sqrt{M_{N}^{\ast 2}+2\nu {q_{p}B%
}}-s\kappa _{p}B\right) ^{2}, \\
{E_{f}^{n}}^{2} &=&k_{f,s}^{n2}+\left( M_{N}^{\ast}-s\kappa _{n}B\right)^{2}.
\end{eqnarray}%
The chemical potentials of nucleons and leptons are given by%
\begin{eqnarray}
\mu_{p} &=&E_{f}^{p}+g_{\omega N}\omega_{0}+g_{\rho N}{\rho_{30}},
\label{eq:mup} \\
\mu_{n} &=&E_{f}^{n}+g_{\omega N}\omega_{0}-g_{\rho N}{\rho_{30}},
\label{eq:mun} \\
\mu_{l} &=&E_{f}^{l}=\sqrt{k_{f,\nu ,s}^{l2}+m_{l}^{2}
                     +2\nu \left\vert {q_{l}}\right\vert B}.  \label{eq:mul}
\end{eqnarray}%
The equation of motion for antikaons
$\bar{K}\equiv \left( K^{-},\bar{K}^{0}\right)$ in neutron star matter with
magnetic fields can be explicitly written as%
\begin{equation}
\left( \partial _{\mu}-iq_{\bar{K}}A_{\mu}-ig_{\omega K}{\omega_{0}}-ig_{\rho
K}\tau_{3K}\rho_{30}\right) \left( \partial ^{\mu}-iq_{\bar{K}}A^{\mu
}-ig_{\omega K}{\omega}_{0}-ig_{\rho K}\tau_{3K}\rho_{30}\right) \bar{K}%
+m_{K}^{\ast 2}\bar{K}=0,  \label{eq:eqk}
\end{equation}%
and the energy spectra are obtained as%
\begin{eqnarray}
\omega_{K^{-}} &=&\sqrt{m_{K}^{\ast 2}+k_{z}^{2}+(2n+1)|q_{K^{-}}|B}-
g_{\omega K}\omega_{0}-g_{\rho K}\rho_{30} ,  \label{eq:ekn} \\
\omega_{\bar{K}^{0}} &=&\sqrt{m_{K}^{\ast 2}+k_{x}^{2}+k_{y}^{2}+k_{z}^{2}}-
g_{\omega K}\omega_{0}+g_{\rho K}\rho_{30} .  \label{eq:ek0}
\end{eqnarray}%
For $s$-wave condensation of negatively charged $K^{-}$
and neutral $\bar{K}^{0}$, we have the chemical potentials
\begin{eqnarray}
\mu_{K^{-}} &=& \omega_{K^{-}}(n=k_z=0)
=\sqrt{m_{K}^{\ast 2}+|q_{K^{-}}|B}-g_{\omega K}\omega_{0}-g_{\rho K}\rho_{30},
\label{eq:mukn} \\
\mu_{\bar{K}^{0}} &=& \omega_{\bar{K}^{0}} (k_x=k_y=k_z=0)
=m_{K}^{\ast}-g_{\omega K}\omega_{0}+g_{\rho K}\rho_{30}.  \label{eq:muk0}
\end{eqnarray}%
The number densities of antikaons are given by%
\begin{eqnarray}
\rho_{K^{-}} &=&2\sqrt{m_{K}^{\ast 2}+|q_{K^{-}}|B}~\bar{K}K,
\label{eq:rhokn} \\
\rho_{\bar{K}^{0}} &=&2m_{K}^{\ast}~\bar{K}K.  \label{eq:rhok0}
\end{eqnarray}

For neutron star matter with uniform distributions, the composition of
matter is determined by the requirements of charge neutrality
and $\beta$-equilibrium conditions. In the present calculation with antikaon
condensation, the $\beta$-equilibrium conditions are expressed by
\begin{eqnarray}
\mu_{n}-\mu_{p} &=&\mu_{K^{-}}=\mu_{e}=\mu_{\mu},  \label{eq:beta1} \\
\mu_{\bar{K}^{0}} &=&0,  \label{eq:beta2}
\end{eqnarray}
and the charge neutrality condition is given by
\begin{equation}
\rho_{v}^{p}=\rho_{K^{-}}+\rho_{v}^{e}+\rho_{v}^{\mu},
\label{eq:charge}
\end{equation}%
with $\rho_{v}^{p}$ and $\rho_{K^{-}}$ given in Eqs.~(\ref{eq:rhovp}) and
(\ref{eq:rhokn}). The vector density of leptons has a similar expression to
that of protons
\begin{equation}
\rho_{v}^{l}=\frac{\left\vert {q_{l}}\right\vert B}{2\pi^{2}}\sum_{\nu}
\sum_{s}k_{f,\nu ,s}^{l}.
\end{equation}%
We solve the coupled Eqs.~(\ref{eq:m1})-(\ref{eq:m3})
and (\ref{eq:beta1})-(\ref{eq:charge}) self-consistently at a given baryon
density, $\rho_{B}=\rho_{v}^{p}+\rho_{v}^{n}$,
in the presence of antikaon condensation and strong magnetic fields.
The total energy density of neutron star matter is given by
\begin{equation}
\varepsilon =\varepsilon _{p}+\varepsilon _{n}+\varepsilon _{\bar{K}%
}+\varepsilon _{l}+\frac{1}{2}m_{\sigma}^{2}\sigma ^{2}+\frac{1}{2}%
m_{\omega}^{2}\omega_{0}^{2}+\frac{1}{2}m_{\rho}^{2}{\rho_{30}^{2}},
\end{equation}%
where the energy densities of nucleons, antikaons, and leptons have the
following forms:%
\begin{eqnarray}
\varepsilon _{p} &=&\frac{{q_{p}B}}{4\pi ^{2}}\sum_{\nu}\sum_{s}\left[
k_{f,\nu ,s}^{p}E_{f}^{p}+\left( \sqrt{M_{N}^{\ast 2}+2\nu {q_{p}B}}-s\kappa
_{p}B\right) ^{2}\ln \left\vert \frac{k_{f,\nu ,s}^{p}+E_{f}^{p}}{\sqrt{%
M_{N}^{\ast 2}+2\nu {q_{p}B}}-s\kappa _{p}B}\right\vert \right] , \\
\varepsilon _{n} &=&\frac{1}{4\pi ^{2}}\sum_{s}\left\{ \frac{1}{2}%
k_{f,s}^{n}E_{f}^{n3}-\frac{2}{3}s\kappa _{n}BE_{f}^{n3}\left( \arcsin \frac{%
M_{N}^{\ast}-s{\kappa _{n}B}}{E_{f}^{n}}-\frac{\pi}{2}\right) -\left(
\frac{s{\kappa _{n}B}}{3}+\frac{M_{N}^{\ast}-s{\kappa _{n}B}}{4}\right)
\right.  \notag \\
&&\times \left. \left[ \left( M_{N}^{\ast}-s{\kappa _{n}B}\right)
k_{f,s}^{n}E_{f}^{n}+\left( M_{N}^{\ast}-s{\kappa _{n}B}\right) ^{3}\ln
\left\vert \frac{k_{f,s}^{n}+E_{f}^{n}}{M_{N}^{\ast}-s{\kappa _{n}B}}%
\right\vert \right] \right\} , \\
\varepsilon _{\bar{K}} &=&\sqrt{m_{K}^{\ast 2}+|q_{K^{-}}|B}\,\rho_{K^{-}}
+m_{K}^{\ast}\,\rho_{\bar{K}^{0}} , \\
\varepsilon _{l} &=&\sum_{l=e,\mu}\sum_{\nu}\sum_{s}\frac{\left\vert {q_{l}%
}\right\vert B}{4\pi ^{2}}\left[ k_{f,\nu ,s}^{l}E_{f}^{l}+\left( m_{l}^{2}{+%
}2\nu \left\vert {q_{l}}\right\vert {B}\right) \ln \left\vert \frac{k_{f,\nu
,s}^{l}+E_{f}^{l}}{\sqrt{m_{l}^{2}{+}2\nu \left\vert {q_{l}}\right\vert {B}}}%
\right\vert \right] .
\end{eqnarray}%
The antikaons do not contribute directly to the pressure because they are in
the $s$-wave condensation. The pressure of the system can be obtained by
\begin{equation}
P=\sum_{i}\mu_{i}\rho_{v}^{i}-\varepsilon =\mu_{n}\rho_{B}-\varepsilon .
\end{equation}%
We note that the contribution from electromagnetic fields to the energy density
and pressure, $\varepsilon_{f}=P_{f}=B^{2}/8\pi$, is not taken into account in
the present calculation. In general, the strong magnetic fields in neutron stars
can produce magnetic forces that play an important role in determining the
structure of the star~\cite{Latt01}.

%%%%%%%%%%%%%%%%%%%%%%%%%%%%%%%%%%%%%%%%%%%%%%%%%%%%%%%%%%%%%%%%%%%%%

\section{ Results and discussion}

\label{sec:3}

In this section, we analyze the properties of neutron star matter with
antikaon condensation in the presence of strong magnetic fields using the
QMC model. The effective masses of nucleons and kaons in the QMC model are
obtained self-consistently at the quark level, which is the main difference
from the RMF model~\cite{WS86}. To show how the results depend on
the models based on different degrees of freedom, we make a systematic
comparison between the QMC model and the RMF model with the TM1 parameter
set~\cite{ST94}. As a nonlinear version of the RMF model, the TM1 model has
been widely used in many studies of nuclear
physics~\cite{Kaon02,NPA98,HS96,PRC02,PTP06}. The TM1 model includes nonlinear
terms for both $\sigma$ and $\omega$ mesons, and its parameters were determined
by reproducing the properties of nuclear matter and finite nuclei including
neutron-rich nuclei~\cite{ST94}. It has be pointed in Refs.~\cite{PBW00,BB01}
that antikaon condensation in the RMF models is quite sensitive to the
antikaon optical potential at normal nuclear matter density,
$U_{\bar{K}}\left( \rho_{0}\right) =g_{\sigma K}\sigma -g_{\omega K}\omega_{0}$.
In the QMC model, the antikaon optical potential is given by
$U_{\bar{K}}=m_{K}^{\ast}-m_{K}-g_{\omega K}\omega_{0}$~\cite{Panda05}, and we
obtain $U_{\bar{K}}\left( \rho_{0}\right) =-123$ MeV with the parameters used
in the present study, which agrees with the result given in Ref.~\cite{Panda05}.
We note that the basic coupling constants in the QMC model are defined at
the quark level. Therefore, $U_{\bar{K}}\left( \rho_{0}\right)$ is a predicted
value with determined $g_{\sigma}^{q}$ and $g_{\omega}^{q}$.
However, $U_{\bar{K}}\left( \rho_{0}\right)$ in the RMF model is
usually taken to be in a range and used to determine the kaon-meson couplings.
Here we take $g_{\omega K}=g_{\omega N}/3$ and $g_{\sigma K}=0.926$ in the TM1
model, which is determined by fitting $U_{\bar{K}}\left(\rho_{0}\right)=-123$
MeV obtained in the QMC model, so the comparison between the QMC and TM1
models is more meaningful.
It is well known that the direct URCA process produces the most powerful
neutrino emission in the core of neutron star and leads to rapid cooling
of the star. In the QMC model at $B=0$, the direct URCA process occurs at the
critical density $\rho_{\mathrm{URCA}} \approx 0.29$ fm$^{-3}$, which implies the
direct URCA process can take place for neutron star with gravitational mass
$M > 1.08\,M_\odot$. These values in the TM1 model are
$\rho_{\mathrm{URCA}} \approx 0.21$ fm$^{-3}$ and $M > 0.81\,M_\odot$.
We note that the maximum masses of neutron stars adopting the QMC and TM1
equations of state with antikaon condensation in the field-free case are
about $1.85\,M_\odot$ and $2.09\,M_\odot$, respectively.

In this article, we present numerical results for the magnetic field strengths
$B^{*}=B/B_{c}^{e}=0$, $10^{5}$, and $10^{6}$
($B_{c}^{e}=4.414\times 10^{13}$ G is the electron critical field). We note
that the magnetic fields $B^*=10^5$ and $10^6$ may be too large for static
configurations of neutron stars to exist. In Ref.~\cite{AA95}, the authors
performed relativistic calculations of axisymmetric neutron star. They
found that the maximum allowable poloidal magnetic field is of the order of
$10^{18}$ G when the magnetic pressure is comparable to the fluid pressure
at the center of the star. In Ref.~\cite{Latt01}, the authors studied
static neutron stars with poloidal magnetic fields and a simple class
of electric current distributions, and they obtained that the maximum magnetic
field strengths at the center of neutron stars
could be as large as about $5\times 10^{18}$ G.

\begin{figure}[htb]
\includegraphics[bb=54 228 478 797, width=8.5 cm,clip]{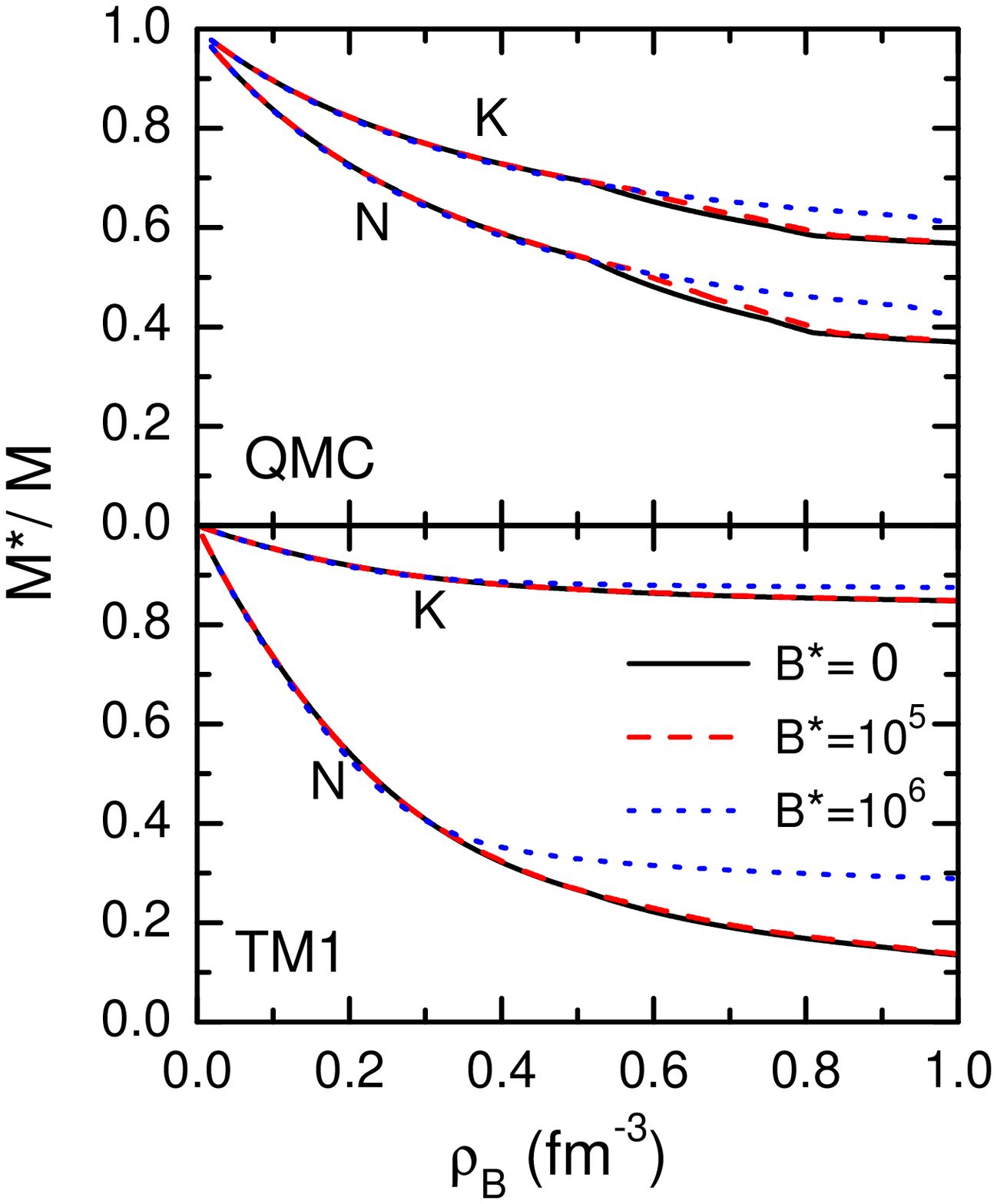}
\caption{(Color online) The effective masses of nucleons and
kaons, $M^{\ast}/M$, as a function of the baryon density, $\rho_{B}$,
for the magnetic field strengths $B^{*}=0$, $10^{5}$, and $10^{6}$.
The results of the QMC model shown in the upper panel are compared with those
of the TM1 model shown in the lower panel.}
\label{fig:rm}
\end{figure}

\begin{figure}[htb]
\includegraphics[bb=47 315 567 803, width=10 cm,clip]{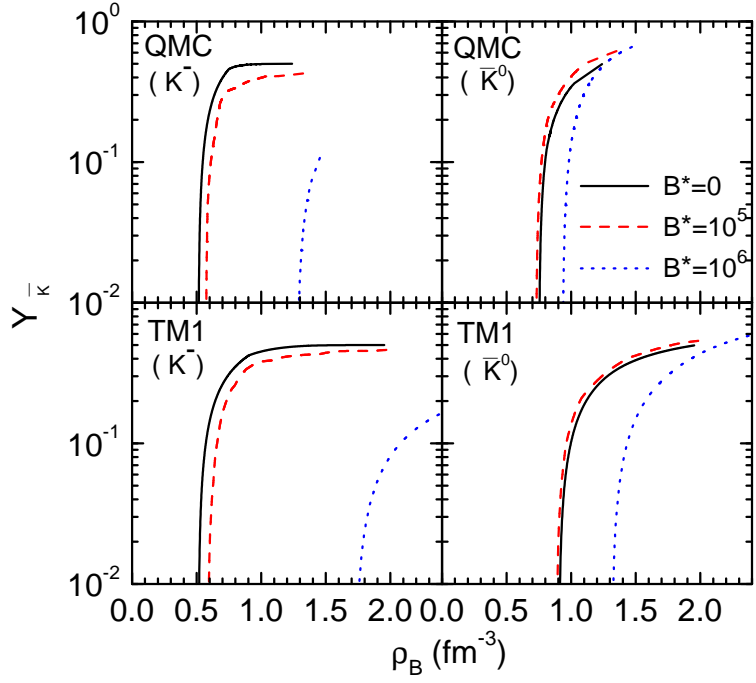}
\caption{(Color online) The antikaon fractions, $Y_{K^{-}}$ (left panels) and
$Y_{\bar{K}^{0}}$ (right panels), as a function of the baryon
density, $\rho_{B}$, for $B^{*}=0$, $10^{5}$, and $10^{6}$.
The results of the QMC and TM1 models are shown in the upper and lower panels,
respectively.}
\label{fig:yk}
\end{figure}

We show in Fig.~\ref{fig:rm} the effective masses of nucleons and kaons as a
function of the baryon density in neutron star matter for
$B^{*}=0$, $10^{5}$, and $10^{6}$. The results of the QMC model
in the upper panel are compared with those of the
TM1 model in the lower panel. It is shown that the effective nucleon masses
in the QMC model are larger than those in the TM1 model, but the opposite
is true for the effective kaon masses. Note that the effective masses of nucleons
and kaons in the QMC model are obtained self-consistently at the quark
level, whereas they are simple linear functions of $\sigma$ in the TM1 model.
We find that the influence of the magnetic field on the effective masses is
not observable until $B^{*}>10^{5}$, and the effective masses of both
nucleons and kaons in strong magnetic fields are larger than the field-free
values. In Fig.~\ref{fig:yk}, we present the antikaon fractions,
$Y_{K^{-}}=\rho_{K^{-}}/\rho_{B}$ (left panels) and
$Y_{\bar{K}^{0}}=\rho_{\bar{K}^{0}}/\rho_{B}$ (right panels), again for
$B^{*}=0$, $10^{5}$, and $10^{6}$. It is seen that the threshold density
of antikaon condensation is significantly changed by the magnetic field
when $B^{*}$ is large enough. In the QMC model (upper panels), the threshold
densities of $K^{-}$ condensation are 0.513,
0.572, and 1.277 $\mathrm{fm}^{-3}$ corresponding to
$B^{*}=0$, $10^{5}$, and $10^{6}$,
whereas those of $\bar{K}^{0}$ condensation are
0.750, 0.728, and 0.934 $\mathrm{fm}^{-3}$,
respectively. In contrast, for the TM1 model in the lower panels,
the threshold densities are 0.514, 0.588,
and 1.732 $\mathrm{fm}^{-3}$ for $K^{-}$
condensation, whereas they are 0.905, 0.885,
and 1.316 $\mathrm{fm}^{-3}$ for $\bar{K}^{0}$ condensation.
We note that the results of the TM1 model shown in Refs.~\cite{Kaon02,PBW00}
can be obtained using twice lower $g_{\rho N}$ than given in Ref.~\cite{ST94}.
Therefore, they are different from our results of the TM1 model.

It is very interesting to compare the effect of the magnetic field on $K^{-}$
and $\bar{K}^{0}$ condensations. The onset of $K^{-}$
condensation shifts to higher density in the presence of strong magnetic
fields, and it even occurs beyond the threshold of $\bar{K}^{0}$
condensation. This is mainly because the negatively charged $K^{-}$ gets a
large chemical potential in the presence of strong magnetic fields, as given in
Eq.~(\ref{eq:mukn}) due to the term $|q_{K^{-}}|B$. However, the
threshold density of $\bar{K}^{0}$ condensation for $B^{*}=10^{5}$ is
slightly smaller than the field-free value, whereas the one for
$B^{*}=10^{6}$ significantly increases. Although $\bar{K}^{0}$ is a neutral
particle, its chemical potential given by Eq.~(\ref{eq:muk0}) is
indirectly influenced by the magnetic field through the dependence of meson
mean fields on the magnetic field. We conclude that $K^{-}$ condensation
depends more on the magnetic field than does $\bar{K}^{0}$ condensation.
By comparing the results of the QMC model with those of the TM1 model,
one can see that the onset of antikaon condensation in the presence of strong
magnetic fields occurs at lower densities within the QMC model. This is because
the chemical potentials of antikaons in the QMC model decrease more rapidly
than those in the TM1 model. We note that the threshold densities of antikaon
condensation is mainly determined by the behavior of the chemical potentials,
and the chemical potentials are related to the meson fields
by Eqs.~(\ref{eq:mukn}) and (\ref{eq:muk0}).
Because the TM1 model includes nonlinear terms for both $\sigma$ and
$\omega$ mesons, the density dependence of the meson fields in the TM1 model
is quite different from those in the QMC model, and it leads to different
behavior of the chemical potentials in the two models. Although there are
quantitative differences between the QMC and TM1 models, we find that
qualitative trends of magnetic field effects are similar in the two models.

\begin{figure}[htb]
\includegraphics[bb=70 349 552 806, width=10 cm,clip]{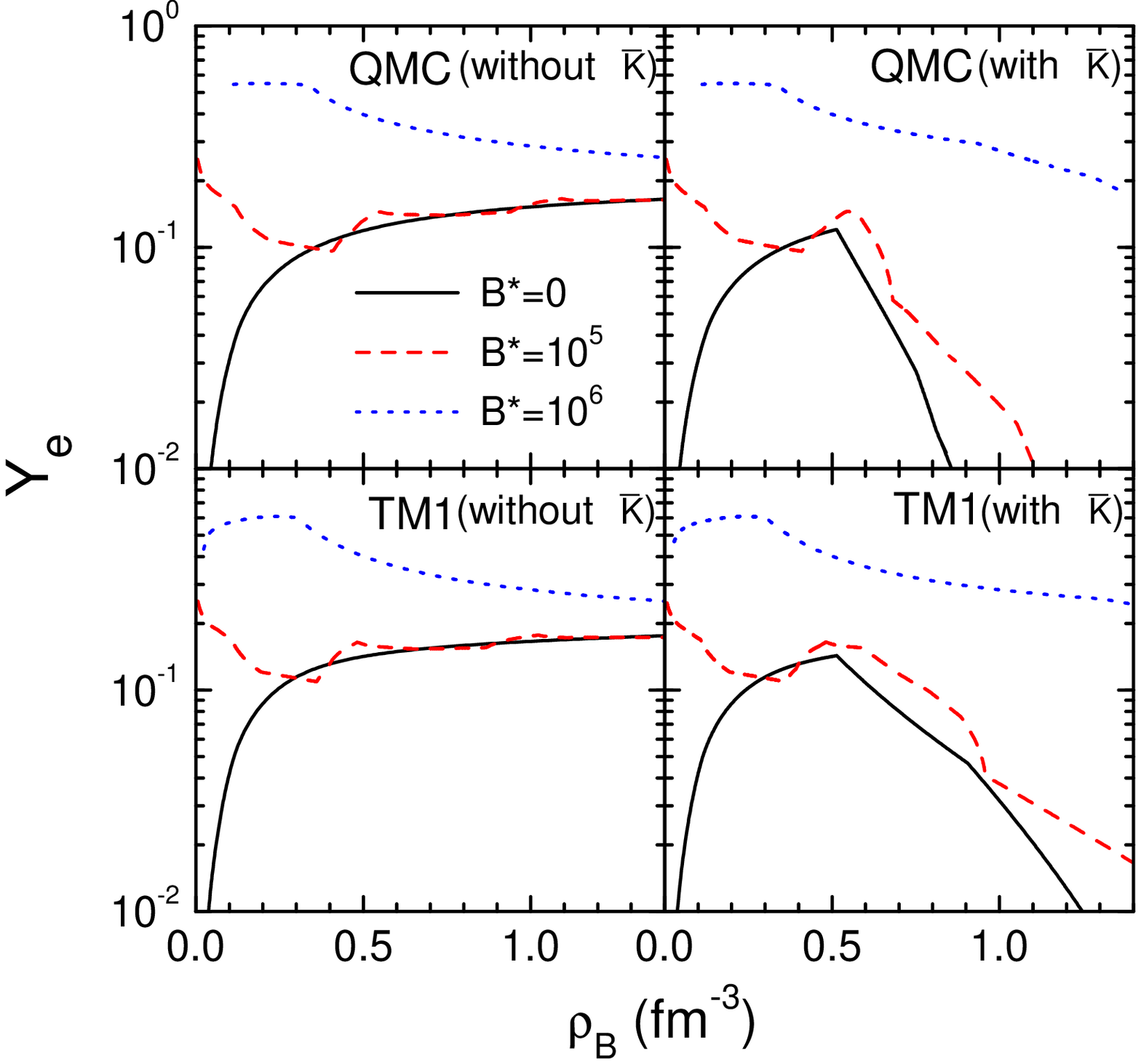}
\caption{(Color online) The electron fraction, $Y_e$, as a function of
the baryon density, $\rho_{B}$, for $B^{*}=0$, $10^{5}$, and $10^{6}$.
The results with and without $\bar{K}$ condensation are shown in the right and left
panels, respectively. The results of the QMC model shown in the upper panels are
compared with those of the TM1 model shown in the lower panels.}
\label{fig:ye}
\end{figure}

\begin{figure}[htb]
\includegraphics[bb=70 349 552 806, width=10 cm,clip]{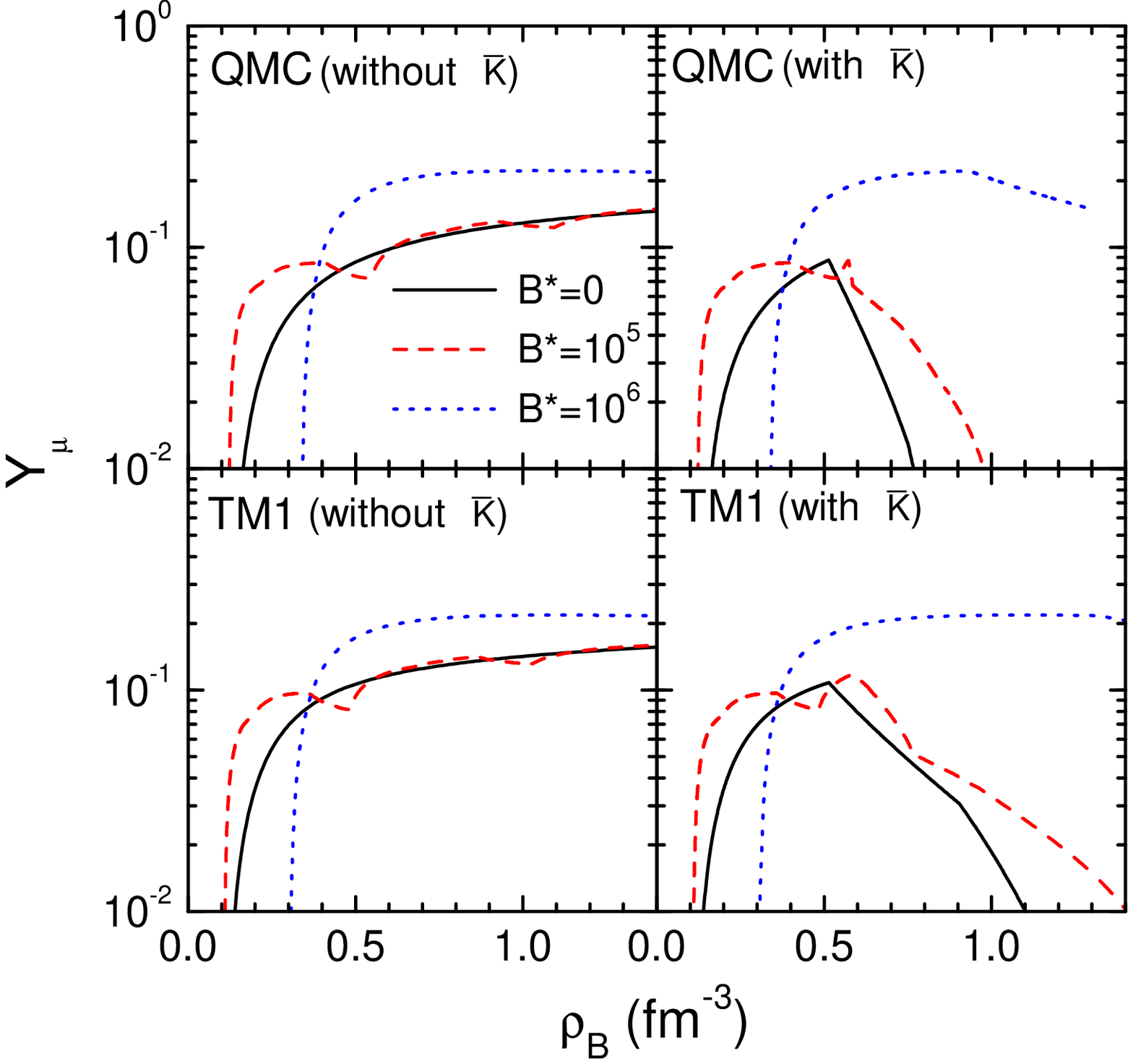}
\caption{(Color online) The muon fraction, $Y_{\mu}$, as a function of
the baryon density, $\rho_{B}$, for $B^{*}=0$, $10^{5}$, and $10^{6}$.
The results with and without $\bar{K}$ condensation are shown in the right and left
panels, respectively. The results of the QMC model shown in the upper panels are
compared with those of the TM1 model shown in the lower panels.}
\label{fig:ym}
\end{figure}

\begin{figure}[htb]
\includegraphics[bb=45 315 572 798, width=10 cm,clip]{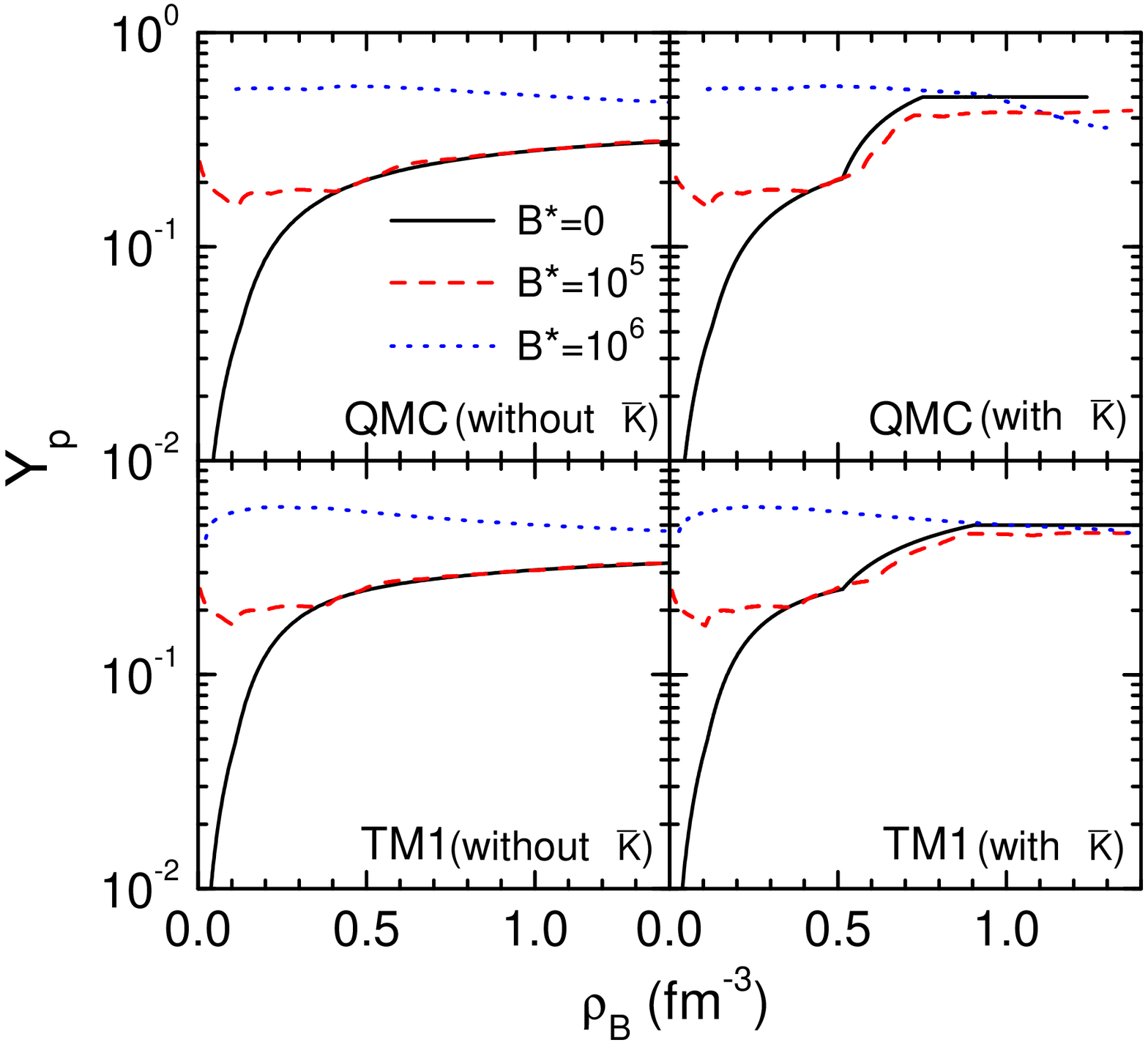}
\caption{(Color online) The proton fraction, $Y_p$, as a function of
the baryon density, $\rho_{B}$, for $B^{*}=0$, $10^{5}$, and $10^{6}$.
The results with and without $\bar{K}$ condensation are shown in the right and left
panels, respectively. The results of the QMC model shown in the upper panels are
compared with those of the TM1 model shown in the lower panels.}
\label{fig:yp}
\end{figure}

In Figs.~\ref{fig:ye}, \ref{fig:ym}, and~\ref{fig:yp}, we show
the electron fraction $Y_{e}=\rho_{v}^{e}/\rho_{B}$,
the muon fraction $Y_{\mu}=\rho_{v}^{\mu}/\rho_{B}$,
and the proton fraction $Y_{p}=\rho_{v}^{p}/\rho_{B}$
as functions of $\rho_{B}$ for $B^{*}=0$, $10^{5}$, and $10^{6}$.
One can obtain the neutron fraction $Y_{n}=\rho_{v}^{n}/\rho_{B}$
as $Y_{n}=1-Y_{p}$. The charge neutrality condition gives the relation
$Y_{p}=Y_{K^{-}}+Y_{e}+Y_{\mu}$.
The results with and without $\bar{K}$ condensation are
displayed in the right and left panels, respectively. The negatively charged
$K^{-}$ play the same role as electrons and muons in maintaining the
charge neutrality. Therefore, the presence of $K^{-}$ condensation
decreases $Y_{e}$ and $Y_{\mu}$ as shown in the right panels
of Figs.~\ref{fig:ye} and \ref{fig:ym}. In the presence of strong magnetic
fields, the chemical potentials $\mu_{e}$, $\mu_{\mu}$, and $\mu_{K^{-}}$
are influenced by the magnetic field. As a result, drops of $Y_{e}$
and $Y_{\mu}$ in Figs.~\ref{fig:ye} and \ref{fig:ym} become less steep
for large magnetic field strength $B^{*}$.
In Fig.~\ref{fig:ym}, it is seen that the threshold densities for the
appearance of muons are affected by the magnetic field.
The threshold of muons for $B^{*}=10^{5}$ is slightly smaller than the
field-free value, whereas the one for $B^{*}=10^{6}$ significantly increases.
The presence of $K^{-}$ condensation leads to an enhancement of $Y_{p}$ as shown
in the right panels of Fig.~\ref{fig:yp}. At high density where $\bar{K}^{0}$
condensation occurs, $Y_{p}=0.5$ is obtained for $B^{*}=0$. This is because
$\beta $-equilibrium conditions enforce $\rho_{v}^{p}=\rho_{v}^{n}$
as discussed in Ref.~\cite{PBW00}. The feature is changed in the presence
of strong magnetic fields, and we get
$E_{f}^{n}-E_{f}^{p}=\sqrt{m_{K}^{\ast 2}+|q_{K^{-}}|B}-m_{K}^{\ast}$
derived from the $\beta $-equilibrium conditions when the condensations
of isospin doublet, $K^{-}$ and $\bar{K}^{0}$,
occur together in neutron star matter. By comparing the right and left
panels of Figs.~\ref{fig:ye}, \ref{fig:ym}, and~\ref{fig:yp}, we find that
the presence of antikaon condensation can significantly change the particle
fractions. It is obvious that the composition of neutron star matter with
antikaon condensation depends on both the magnetic field strength $B^{*}$
and the baryon density $\rho_{B}$. We note that the tendencies in the upper
panels obtained in the QMC model are quite similar to those of the TM1 model
in the lower panels of Figs.~\ref{fig:ye}, \ref{fig:ym}, and~\ref{fig:yp}.
In the right panels of Figs.~\ref{fig:ye} and \ref{fig:ym} where antikaons
are included, $Y_e$ and $Y_\mu$ decrease more rapidly at high densities in
the QMC model than in the TM1 model. This is due to the same reason as
that discussed for the earlier appearance of antikaon condensation in the
QMC model. Because the $\beta$-equilibrium condition
$\mu_{K^{-}}=\mu_{e}=\mu_{\mu}$ should be satisfied,
$\mu_{e}$ and $\mu_{\mu}$ in the QMC model decrease more rapidly
than those in the TM1 model, just like the case of $\mu_{K^{-}}$.
This leads to a quicker drop in the upper right panels of
Figs.~\ref{fig:ye} and \ref{fig:ym} comparing with those in the lower
right panels.

\begin{figure}[htb]
\includegraphics[bb=46 338 563 797, width=10 cm,clip]{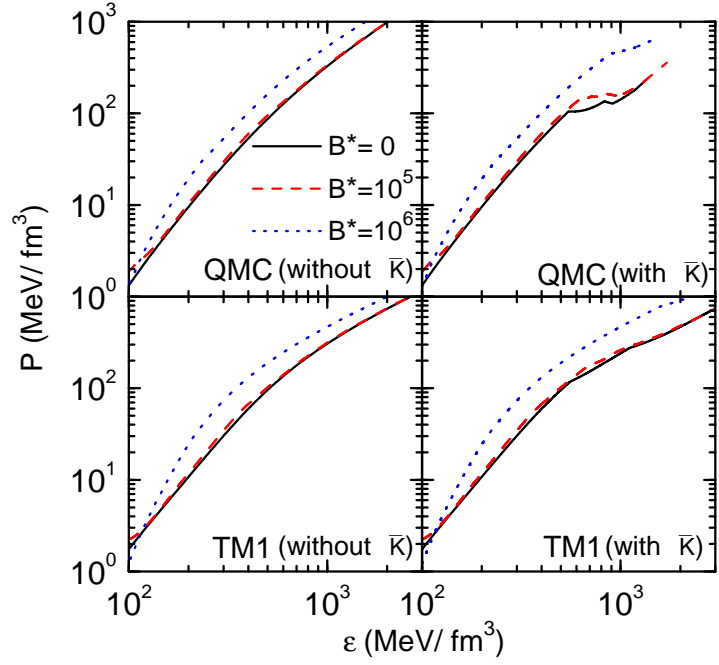}
\caption{(Color online) The matter pressure, $P$, versus the matter
energy density, $\varepsilon$, for $B^{*}=0$, $10^{5}$, and $10^{6}$.
The results of the QMC model with and without $\bar{K}$ condensation
are shown in the upper right and upper left panels, respectively.
Those in the TM1 model are shown in the lower panels for comparison.}
\label{fig:ep}
\end{figure}

In Fig.~\ref{fig:ep}, we show the matter pressure $P$ as a function of the
matter energy density ${\varepsilon}$ for the magnetic field strengths
$B^{*}=0$, $10^{5}$, and $10^{6}$. The results with and without $\bar{K}$
condensation are plotted in the right and left panels, respectively. It is
seen that the presence of $\bar{K}$ condensation makes the EOS softer
compared with the case without $\bar{K}$ condensation. The softening of the
EOS becomes less pronounced with increasing magnetic field for
$B^{*}>10^{5}$. This is because the threshold of antikaon condensation shifts
to higher density and the effect of antikaon condensation on the EOS gets
weaker with increasing $B^{*}$. At higher densities, the influence of
strong magnetic fields on the EOS becomes noticeable as the field
strength increases above $B^{*}\sim 10^{5}$. Here we include the
anomalous magnetic moments of nucleons, which play an important role in the
study of neutron star matter with strong magnetic fields as discussed
in our previous work~\cite{YS06}. The results of the QMC model in the upper
panels are compared with those of the TM1 model in the lower panels.
The EOS in the QMC model is slightly softer than the one in the TM1 model.

%%%%%%%%%%%%%%%%%%%%%%%%%%%%%%%%%%%%%%%%%%%%%%%%%%%%%%%%%%%%%%%%%%%%%

\section{Summary}

\label{sec:4}

In this article, we have studied the effects of strong magnetic fields on
antikaon condensation in neutron star matter using the QMC model. Nucleons
and antikaons in the QMC model are described as MIT bags that interact
through the self-consistent exchange of scalar and vector mesons in the
mean-field approximation. In the QMC model, the effective masses of nucleons
and kaons are obtained self-consistently at the quark level, which is the
main difference from the RMF models. It is clear that the effects of strong
magnetic fields become significant only for magnetic field strength
$B^{*}>10^{5}$. We found that the presence of strong magnetic fields
significantly alters the threshold density of antikaon condensation.
The threshold of antikaon condensation shifts to higher density in the
presence of strong magnetic fields. For $B^{*}=0$, $10^{5}$, and $10^{6}$,
we obtained the threshold densities of $K^{-}$ condensation are
0.513, 0.572, and 1.277 $\mathrm{fm}^{-3}$,
whereas those of $\bar{K}^{0}$ condensation are 0.750,
0.728, and 0.934 $\mathrm{fm}^{-3}$.
It is obvious that the threshold density of $K^{-}$ condensation
depends strongly on the magnetic field strength, and it even shifts beyond
the threshold of $\bar{K}^{0}$ condensation for sufficiently strong
magnetic fields. Because the negatively charged $K^{-}$ can play the same
role as electrons and muons in maintaining the charge neutrality,
the presence of $K^{-}$ condensation decreases the fractions of electrons
and muons dramatically, and the drops become less steep for larger
field strength. It is found that the presence of antikaon condensation
can significantly change the composition of neutron star matter, and the
particle populations depend on both the magnetic field strength $B^{*}$
and the baryon density $\rho_{B}$.

The presence of antikaon condensation can make the EOS softer compared
with the case without antikaon condensation. The softening of the EOS
becomes less pronounced with increasing magnetic field for $B^{*}>10^{5}$.
It is known that the Landau quantization of charged particles causes a
softening in the EOS, whereas the inclusion of nucleon anomalous magnetic
moments leads to a stiffening of the EOS. At high densities, the softening
of the EOS from Landau quantization with increasing magnetic field for
$B^{*}>10^{5}$ can be overwhelmed by the stiffening
resulting from the incorporation of anomalous magnetic moments.
We have made a systematic comparison between the results of the QMC and TM1
models. It is found that quantitative differences exist between the two
models, but qualitative trends of magnetic field effects on antikaon
condensation and EOS are quite similar.

Finally, we would like to give a few remarks on the limitations of the results
presented in this article. Although the QMC model is constructed at the quark
level, it is certainly approximate due to several reasons. The MIT bag model
itself is known to be a simplification, and the mean-field treatment for
meson fields is an approximation. The adopted model of spherical bags can
be violated in a superstrong magnetic field, because the magnetic localization
length for protons on the ground Landau level with $B^{*} \sim 10^{6}$
is of the same order as the bag radius used in the present calculation.
These approximations need to be examined more carefully in further work.
For neutron star matter at high density, hyperons may appear before
the onset of antikaon condensation. It would be interesting and important
to include hyperons and antikaon condensation in the study of neutron star
matter with strong magnetic fields.

%%%%%%%%%%%%%%%%%%%%%%%%%%%%%%%%%%%%%%%%%%%%%%%%%%%%%%%%%%%%%%%%%%%%%

\section*{Acknowledgments}

This work was supported in part by the National Natural Science Foundation
of China (No. 10675064) and the Specialized Research Fund for the Doctoral
Program of Higher Education (No. 20040055010).

%%%%%%%%%%%%%%%%%%%%%%%%%%%%%%%%%%%%%%%%%%%%%%%%%%%%%%%%%%%%%%%%%%%%%

%%%%%%%%%%%%%%%%%%%%%%%%%%%%%%%%%%%%%%%%%%%%%%%%%%%%%%%%%%%%%%%%%%%%%
\newpage

%%%%%%%%%%%%%%%%%%%
%\newpage

%%%%%%%%%%%%%%%%%%%
%\newpage

%%%%%%%%%%%%%%%%%%%
%\newpage

%%%%%%%%%%%%%%%%%%%
%\newpage

%%%%%%%%%%%%%%%%%%%
%\newpage

%%%%%%%%%%%%%%%%%%%%%%%%%%%%%%%%%%%%%%%%%%%%%%%%%%%%%%%%%%%%%%%%%%%%%%%%%%
\end{document}